\documentclass[final]{svjour3}
\usepackage{graphicx}
\usepackage{rotating}
\usepackage{amssymb}
\usepackage{mathptmx}
\usepackage[numbers]{natbib}
\makeatletter
\journalname{Journal of Low Temperature Physics}

\newcommand{\HeThree}{$^3$He}
\newcommand{\HeFour}{$^4$He}

\bibpunct{}{}{,}{s}{}{,}

\begin{document}

\newcommand{\hdblarrow}{H\makebox[0.9ex][l]{$\downdownarrows$}-}

\title{The glassy response  of solid $^4$He to torsional oscillations}

\author{M. J. Graf$^1$ \and Z. Nussinov$^2$ \and A. V. Balatsky$^{1,3}$}

\institute{1: Theoretical Division, Los Alamos National Laboratory, Los Alamos, New Mexico 87545, USA
\\2: Department of Physics, Washington University, St.\ Louis, Missouri 63160, USA
\\3: CINT, Los Alamos National Laboratory, Los Alamos, New Mexico 87545, USA}

\date{07.23.2009}

\maketitle

\begin{abstract}
We calculated the glassy response of solid \HeFour\ to torsional oscillations
assuming a phenomenological glass model.
Making only a few assumptions about the distribution of glassy relaxation times in a small subsystem of
otherwise rigid solid \HeFour, we can account for the magnitude of the observed period shift and
concomitant dissipation peak in several torsion oscillator experiments.
The implications of the glass model for solid \HeFour\ are threefold:
(1) The dynamics of solid \HeFour\ is governed
by glassy relaxation processes.
(2) The distribution of relaxation times varies significantly between
different torsion oscillator experiments.
(3) The mechanical response of a torsion oscillator does not require a
supersolid component to account for the observed anomaly at low temperatures,
though we cannot rule out its existence.
\end{abstract}


\section{Introduction}
Torsion oscillators have been used successfully to measure an anomalous change in resonant period
and accompanying dissipation in solid
\HeFour.\cite{Chan04, Rittner06, Kondo07, Aoki07, Clark07, Penzev07, Hunt09}
It has been speculated that the observed signature is due to Bose-Einstein condensation of vacancies or interstitials
forming a novel supersolid state in otherwise crystalline
\HeFour.\cite{Andreev69, Chester67, Reatto69, Chester70, Leggett70, Anderson84}
Early on, the change in resonant period has been attributed to a (nonclassical)
decoupling of a supersolid component from the (classical) normal moment of inertia.
This is not surprising, since the observed change in period is in agreement
with similar observations of onset of superfluidity in liquid \HeFour, measured a long time ago by torsion
oscillators.\cite{Andronikashvili46, Andronikashvili66, HF,  London}
However, it is important to remember that for liquid \HeFour\ it was already well established,
long before the torsion oscillator (TO) experiments were performed,
that it undergoes a  transition from liquid to superfluid with no viscosity.
It was natural to use the connection between superflow and period drop.  Status
of a search of supersolidity in  solid \HeFour \ is different: a change in period has been reported, but no evidence of mass
superflow\cite{Greywall77, Beamish05, Beamish06, Sasaki06, Ray08, Bonfait89, Balibar08} or
condensation \cite{Diallo07, Blackburn07, Kirichek08} has been seen below the expected transition temperature.
It is therefore necessary to ask the question what is the relationship between the change in period and
superflow, and what alternate physical mechanisms can explain the change in period and concomitant peak in dissipation.

\begin{figure}
\bigskip
\begin{center}
\includegraphics[width=0.62\linewidth,angle=0,keepaspectratio]{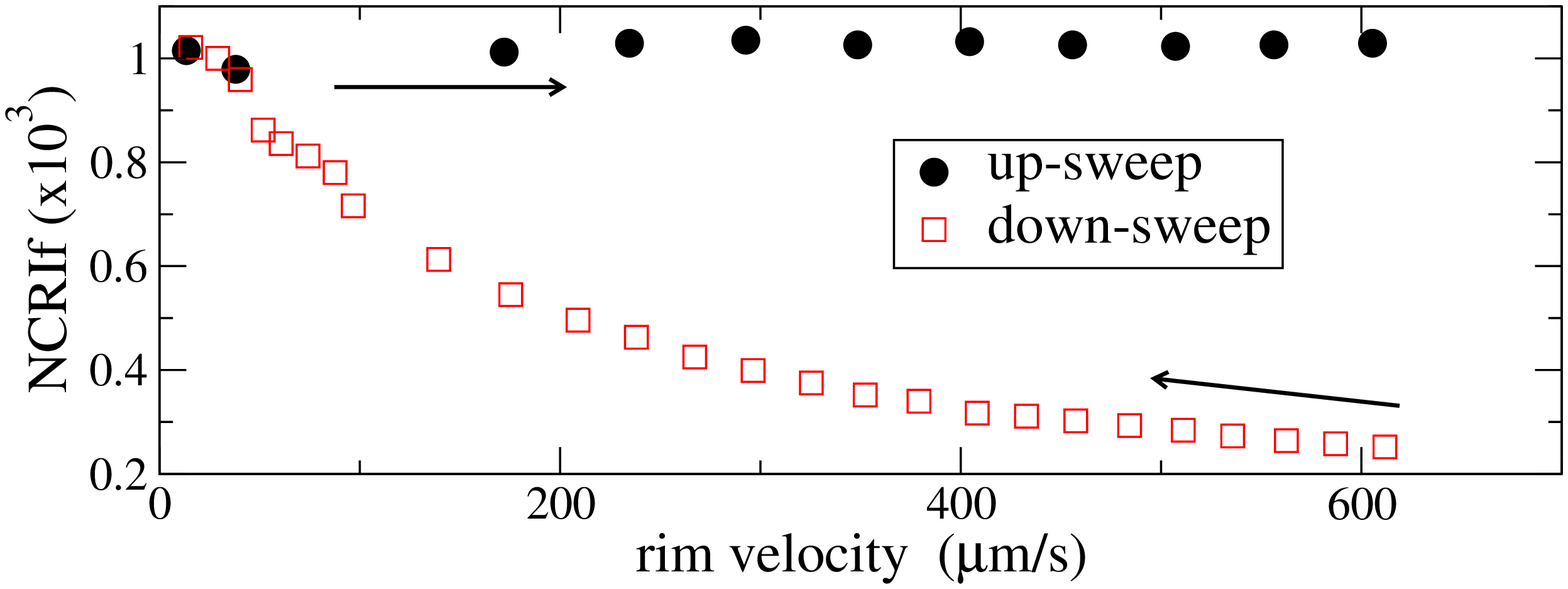}
\end{center}
\caption{(Color online) Hysteresis of rim velocity of nonclassical rotational inertia fraction (NCRIf)
by Aoki et al.\cite{Aoki07} Measurements were taken at 19 mK as the oscillator drive was
increased (circle) and decreased (square).
}\label{Fig0}
\end{figure}

A potential contender for an alternate explanation is the glass scenario, which automatically accounts for hysteresis effects,
annealing dependence, linear term in specific heat, and long relaxation times in many observables.
For example, the strong hysteresis effect reported by Aoki et al.\cite{Aoki07} for the rim velocity dependence of the torsion
oscillator, shown in Fig.~\ref{Fig0}, is consistent with a glassy response.
However, it is difficult to explain within a supersolid scenario how increasing the rim velocity of the cell
 does not change the reported nonclassical rotational inertia: between slow and fast rim velocity one increases kinetic energy of solid, $E_{kin} \propto v^2$, and the ratio of kinetic energies is on the  order of $\sim (600/10)^2 = 3600$. Fast rim velocity  exceeds the supposed critical velocity of the condensate, $\sim 10 \, {\rm \mu m/s}$, by
several orders of magnitude.

In this article, we explore if a phenomenological glass model can account for the change in period without
postulating a nonclassical moment of inertia. Similar to the supersolid picture, we assume that only a small subsystem
of solid \HeFour\ exhibits glassy properties that dominate the response at low temperatures.
This is an important point, since it has been argued before that the observed large change in dissipation cannot be described
by uniform Bose-Einstein condensation.\cite{Nussinov07, Graf08, Huse07}
It remains to be seen if nonuniform Bose-Einstein condensation alone either along grain
boundaries \cite{Clark06}
or along the axis of screw dislocations \cite{Boninsegni06, Pollet07, Boninsegni07}
can explain the dynamic response of TOs.
A discussion on the role of a glassy component does not rule in or out the presence of a supersolid component.
We emphasize that our analysis addresses a glassy contribution regardless of the magnitude of a supersolid component.

\section{Glass model for torsion oscillator}

{\em Glass model:} \
In previous work, we argued for the possibility of a glass phase at low temperatures, roughly below $\sim 150$ mK,
to explain the observed anomalous linear temperature dependence in the specific heat of the otherwise perfect Debye solid
\HeFour.\cite{Balatsky07, Graf08}
Below we will give  an extended oscillator analysis for which the exact nature  of the glass is  not crucial.
For example, it may be caused by two-level systems (TLS) of pinned dislocation lines, vortex excitations, etc.
However, it is important to point out that the amplitude of the period shift and dissipation peak
can be changed dramatically by the growth history or annealing process of the crystal.
In order to explain these puzzling features of solid \HeFour, we conjectured earlier\cite{Balatsky07} that structural defects, e.g.,
localized dislocation segments, form a set of TLS observable at low temperatures.
These immobile crystal defects affect the {\em thermodynamics}\cite{Balatsky07,Graf08}  of bulk \HeFour\
and the {\em mechanics} \cite{Nussinov07} of the TO loaded with \HeFour.
For the analysis of the specific heat, we used independent TLS to obtain the universal glass signature of a
linear-in-temperature specific heat term at low temperatures.
In parallel, we used a phenomenological glass  model, that may originate from an ensemble of TLS,
to describe the mechanical response of the TO.

{\em Torsion oscillator and rotational susceptibility:} \
To set the discussion, we note that TO experiments measure the period and dissipation.
One applies a force and generates a displacement of the oscillator.
The relationship between the force and displacement (angle) is controlled by the TO susceptibility.
In order to extract information from such an experiment, one needs a model to determine the relation between observable period
and dissipation of the TO and the corresponding moment of inertia, damping and effective stiffness of the media.
We start with the general equation of motion
for a harmonic TO defined by an angular coordinate
$\theta$ in the presence of an external and internal torque,\cite{Nussinov07}
\begin{eqnarray}
I_{osc} \ddot{\theta}(t) + \gamma_{osc} \dot{\theta}(t) + \alpha_{osc} \theta(t)
= {M}_{ext}(t) + {M}_{int}(t).
\label{de}
\end{eqnarray}
Here, $I_{osc}$ is the moment of inertia of the (empty) TO chassis,
$\alpha_{osc}$ is the restoring (stiffness) coefficient of the torsion rod, and $\gamma_{osc}$ is its
dissipative coefficient. $M_{ext}(t)$ is the externally imposed torque by the drive.
${M}_{int}(t) = \int {g}(t-t') \theta(t') dt'$
is the internal torque exerted by solid \HeFour\ on the oscillator for a system with time
translation invariance.  In general, the backaction $g(t-t')$ is temperature, $T$, dependent.
The experimentally measured quantity is the angular motion of the TO - {\em not} that of bulk helium, which is enclosed in it. Ab
initio, we cannot assume that the medium moves as one rigid body.
If the non-solid subsystem ``freezes" into a glass, the medium will move with greater
uniformity and speed. This leads to an effect similar to that of the nonclassical
rotational moment of inertia, although its physical origin is completely different.
Therefore, we argue for an alternate physical picture, namely that of softening of the oscillator's stiffness.
The angular coordinate $\theta(t)$ of the TO is a convolution of the applied
external torque ${M}_{ext}(t)$ with the TO susceptibility ${\chi}(t,t')$.
Under Fourier transformation the angular response of the TO is
\begin{eqnarray}
\chi_0^{-1}(\omega)\theta(\omega) =  M_{ext}(\omega) + M_{int}(\omega) .
\label{ft}
\end{eqnarray}
Defining the total angular susceptibility as
$\chi^{-1} = \chi_0^{-1} - M_{int}$,
we write
\begin{eqnarray}
\chi^{-1}(\omega) =
\alpha_{osc} - i \gamma_{osc} \omega - I_{osc} \omega^{2} - g(\omega),
\label{central}
\end{eqnarray}
where $g(\omega)$ is the Fourier transform of the backaction
due to the added solid \HeFour. In what follows, we will treat the
backaction as a small perturbation to the TO chassis.

\section{Period and dissipation of torsion oscillator}
We now determine the experimental consequences of the phenomenological glass model,
where a small glassy subsystem of solid \HeFour\ gives rise to the observed dynamic behavior.
Glass is generally defined as a supercooled liquid out of equilibrium on measurable
time scales: its equilibration time becomes extremely large
(and unmeasurable) at low temperatures. Any glass former
is a liquid at high temperature and becomes an amorphous solid (the glass)
at sufficiently low temperatures. In this context bulk solid \HeFour \ is not a glass former;
we are talking about glass forming within a small fraction of \HeFour \ sample. We note that our analysis will
remain qualitatively unchanged for a general description of the system
by a "freezing" at low temperature of the appropriate component (defect or other)
that is dynamic at high temperatures (see the appendix of Nussinov et al.\cite{Nussinov07}).

We start by reviewing results for an underdamped harmonic torsion oscillator.
The resonant period is obtained from the angular coordinate
$\theta(t) = {\rm Re}\{\theta_0 \exp[-i \omega t] \}$,
with a complex amplitude $\theta_0$ and complex frequency
$\omega = \omega_{0} - i \kappa$.
In the case of an underdamped TO with $\kappa \ll \omega_0$, the resonant period is
$P = {2 \pi}/{\omega_{0}}$,
and the quality factor $Q$ or dissipation is
$Q^{-1} = {2\kappa}/{\omega_{0}}$,
with resonant frequency $\omega_0 = \sqrt{\alpha_{osc}/I_{osc}}$.

In the remainder, we use effective oscillator parameters, which are defined as
the sum of parameters describing the chassis, $\chi_0^{-1}$,
and the added solid \HeFour\ given by
\begin{equation}
g(\omega)=i\gamma_{He}\omega + I_{He}\omega^2 + {\cal G}(\omega) .
\label{geq}
\end{equation}
Thus, we write the net moment of inertia $I = I_{osc} + I_{He}$
and dissipation $\gamma=\gamma_{osc}+\gamma_{He}$.
The total response function of the TO is given by Eqns.~(\ref{central}) and (\ref{geq}).
The term ${\cal G}(\omega)$ captures the dynamics of a glass component and is a function of temperature and frequency.
In the limit $\omega \to 0$ the term ${\cal G}(\omega) \to 0$ as the mechanical motion of any glass component will be the same
as that of the surrounding solid. Hence there will be no relative motion and no transient overdamped modes for $\omega=0$.
However, at any finite frequency $\omega$, we can approximate the glass response by
${\cal G}(\omega) \approx g_0 G(\omega)$,
where the coefficient $g_0$ measures the glassy contribution of the solid and is evaluated at the resonant
frequency $\omega_0$ of the TO.
The dynamic response function $G(\omega)$ of a glass can be approximated by a distribution of
overdamped oscillators with different relaxation times $\tau$. Two popular
relaxation time distributions used
in the literature are the Cole-Cole (CC) and Davidson-Cole (DC) functions. Both describe a
superposition of overdamped oscillators.\cite{Phase1, Phase2}
The CC distribution  gives
$G(\omega)  = 1/[ 1- (i \omega \tau)^{\alpha}]$, while the DC distribution results in
$G(\omega)  = 1/[ 1- i \omega \tau]^{\beta}$.

By comparison to Eq.~(\ref{central}) for the TO chassis system with no helium, the
glassy part of the backaction of \HeFour, ${\cal G}(\omega)$,
renormalizes the effective spring stiffness \cite{Nussinov07, Dorsey08}
\begin{eqnarray}
\alpha^{eff} \simeq (\alpha_{osc} - g_{0}), ~~~ \mbox{for} ~ \omega \tau \ll 1,
\label{eq_shift1}\\
\alpha^{eff} \simeq  \alpha_{osc}, ~~~~ \mbox{for} ~ \omega \tau \gg 1.
\label{eq_shift2}
\end{eqnarray}
These expressions flesh out the dependence of the medium response on the applied driving frequency.
When the driving frequency is far more rapid, $\omega  \gg \tau^{-1}$, then
the transient response of the medium is that of a liquid.
In that limit, the transient modes within the medium cannot ``keep up'' with
the driving torque and only the bare stiffness of the TO remains augmented by the
solid helium contribution.  The effective spring stiffness is that of the driving oscillator,
$\alpha^{eff} = \alpha_{osc}$, see Eq.~(\ref{eq_shift2}).
The limit $\tau^{-1} \to 0$ corresponds to that
of an ideal rigid low-temperature glass in which no transient liquid-like
response of the system is present.
By contrast,  for slow oscillations
$\omega \ll \tau^{-1}$, the excited modes within \HeFour\ are of characteristic transient
time $\tau$ that is long enough to respond to the
driving torque and lead to an additional backaction and
effective reduction of the spring stiffness, see Eq.~(\ref{eq_shift1}).
From this discussion it is clear that the maximum relative shift in period or frequency
will depend on the glassy fraction $g_0$ given by
$\Delta\omega_{max}/\omega_0 \sim g_0/(2 \alpha_{osc})$,
which can vary widely between different torsion oscillators, growth and annealing procedures.

The resonant frequency of the TO with backaction is given by the root of
\begin{eqnarray}
\chi^{-1}(\omega) =
\alpha - i \gamma \omega - I \omega^{2}- g_{0} G(\omega) \equiv 0.
\label{central_glass}
\end{eqnarray}
We anticipate that when the relaxation time is similar to the period of the
underdamped TO, the dissipation will be maximal.
Here, the glassy component responds with the same frequency as the ``normal'' solid component.
The glassy part merely renormalizes the effective spring constant $\alpha$,
but does not lead to additional transient modes, which closely interfere with
the oscillations of the ``normal'' part of the TO.
We look for the largest magnitude of the imaginary part of the root
and see when it is maximal as a function of $\tau$. A larger imaginary
part implies a shorter decay time and a smaller value of $Q^{-1}$.
Since the homogeneous Eq.~(\ref{central_glass}) is scale invariant,
we normalize all oscillator quantities by the effective moment of inertia $I$, i.e.,
$\bar{\alpha} = \alpha/I$, $\bar{\gamma} = \gamma/I$, and $\bar{g}_0 = g_0/I$.

As can be seen from Eq.~(\ref{central_glass}), for an ideal dissipationless oscillator, $\bar\gamma=0$,
the resonant frequency
$\omega_{0}= \sqrt{\bar\alpha}$
is the pole of $\chi(\omega)$ in the limit $\tau^{-1} \to 0$. If we expand $\chi^{-1}$ about this root,
$\omega= \omega_{0} + \delta\omega$, with $\delta\omega = \omega_{1} - i \kappa$,
then we find to leading order in $\delta\omega$
\begin{eqnarray}
\delta\omega \approx
-\frac{i \bar\gamma \omega_{0} +
\bar{g}_0 G(\omega_0)
}{i \bar\gamma + 2\omega_{0}} .
\end{eqnarray}
Therefore, the root attains an imaginary component $\kappa$ and the dissipation becomes
\begin{eqnarray}
Q^{-1} = \frac{2 \kappa}{\omega_{0}}
\approx \frac{A}{\omega_{0}} {\rm Im\ } G(\omega_0) + Q_\infty^{-1} ,
\label{Q-1p}
\end{eqnarray}
with $A = \bar{g}_0/\omega_{0}$ and  $Q_{\infty}^{-1} = {\bar\gamma}/{\omega_{0}}$.
As $\omega_0$ increases for fixed $\alpha_{osc}$, $Q_{\infty}^{-1}$ increases.
For $\alpha\simeq\beta\simeq 1$, the dissipation peaks near $\omega_{0}\tau \sim 1$.
Similarly the resonant frequency becomes
\begin{eqnarray}
2\pi f \equiv \frac{2\pi}{P}
&\approx& \omega_{0} - \frac{A}{4 \omega_{0}}
\Big(
  2 \omega_{0} \, {\rm Re\ }G(\omega_0) + \bar\gamma \, {\rm Im\ }G(\omega_0)
\Big),
\label{P-1p}
\end{eqnarray}
which increases monotonically when $T$ is lowered.
For the special case of Debye relaxation processes ( $\alpha = \beta=1$), we find
${\rm Re\ }G(\omega_0) = [ 1 + (\omega_{0}\tau)^2 ]^{-1}$ and
${\rm Im\ }G(\omega_0) = \omega_{0}\tau[ 1 + (\omega_{0}\tau)^2 ]^{-1}$ and
recover results reported
earlier,\cite{Nussinov07} except for the
additional contribution proportional to  $\bar{\gamma}$ in Eq.~(\ref{P-1p}). It
follows that the changes in dissipation, $\Delta Q^{-1} = Q^{-1} - Q^{-1}_\infty$,
and frequency, $\Delta \omega = \omega_{0} - 2\pi/P$, determine the glass relaxation time $\tau$.
Combining Eqns.~(\ref{Q-1p}) and (\ref{P-1p}) we arrive at a general relation between shift in
dissipation and frequency for $\bar{\gamma}\tau \ll 1$:
\begin{equation}\label{Q-P-ratio}
\frac{\Delta Q^{-1}}{\Delta \omega} =
\frac{ 4 {\rm Im\ }G }{ 2\omega_0{\rm Re\ }G + \bar{\gamma} {\rm Im\ }G } \approx
\frac{2}{\omega_{0}} \frac{ {\rm Im\ }G }{ {\rm Re\ }G }.
\end{equation}
For example, for a DC glass distribution this becomes
\begin{equation}\label{Q-P-ratio-DC}
\frac{\Delta Q^{-1}}{\Delta \omega} \approx \frac{2}{\omega_{0}}
\tan\big(\beta\, {\rm arctan}(\omega_{0}\tau)\big)
\sim
\left\{
\begin{array}{ll}
2 \beta \tau, & \omega_0\tau \to 0, \\
\frac{2}{\omega_0} \tan( \beta \pi/4 ), & \omega_0\tau = 1, \\
\frac{2}{\omega_0} \tan( \beta \pi/2 ), & \omega_0\tau \to \infty.
\end{array}
\right.
\end{equation}
In the past, there have been several reports of large experimental ratios
${\Delta Q^{-1}}\frac{\omega_0}{\Delta \omega} \sim 3-12$.\cite{Rittner06,Aoki07,Clark07}
Because of $\beta \leq 1$ and for $\omega_0\tau \sim 1$ the ratio is limited to
${\Delta Q^{-1}}\frac{\omega_0}{\Delta \omega} \leq 2$,
this requires diverging relaxation times close to the temperature
where the dissipation peaks. For such cases,
$\omega_0\tau \to \infty$ in Eq.~(\ref{Q-P-ratio-DC}) and ratios of order 10
can be obtained for values of $\beta\sim 0.6$.
On the other hand, for $\beta=1$, Eq.~(\ref{Q-P-ratio}) simplifies even further with
${\Delta Q^{-1}}\approx 2\omega_{0}\tau ({\Delta \omega}/\omega_{0}) $.
Similar results for the ratio were obtained for other phenomenological models with
dissipative channels.\cite{Huse07, Dorsey08}
For example, Huse and Khandker \cite{Huse07} assumed a simple phenomenological
two-fluid model, where the supersolid is dissipatively
coupled to a normal solid resulting in a ratio of
${\Delta Q^{-1}}\frac{\omega_0}{\Delta \omega} \approx 1$. Yoo and Dorsey\cite{Dorsey08}
developed a viscoelastic model and Korshunov\cite{Korshunov09} derived a TLS glass model
for solid \HeFour\ that captures
the results of the general phenomenological glass model originally proposed by Nussinov et al.\cite{Nussinov07}
Here, we like to emphasize that it is challenging to reconcile a large dissipative $\Delta Q/Q$ ratio
with uniform Bose-Einstein condensation.\cite{Nussinov07, Graf08, Huse07}

We now make further assumptions about the glassy
relaxation time $\tau$. In many glass formers $\tau$ follows the phenomenological Vogel-Fulcher-Tamman (VFT) expression
$\tau(T) = \tau_{0} \exp[ D T_{0}/(T-T_{0})]$ for $T>T_{0}$.
Here, $T_{0}$ is the temperature at which an ideal glass transition occurs, which is below
the temperature where the peak in dissipation occurs. The parameter $D$ is a measure of the fragility of
the glass ($D \lesssim 10$ for fragile glasses \cite{Angell95, Borrego02}).
Finally, at temperatures $T<T_{0}$ the glassy subsystem freezes out and
$\tau$ becomes infinite.

\section{Results and discussion}

\begin{figure}
\bigskip
\begin{center}
\includegraphics[width=0.65\linewidth,angle=0,keepaspectratio]{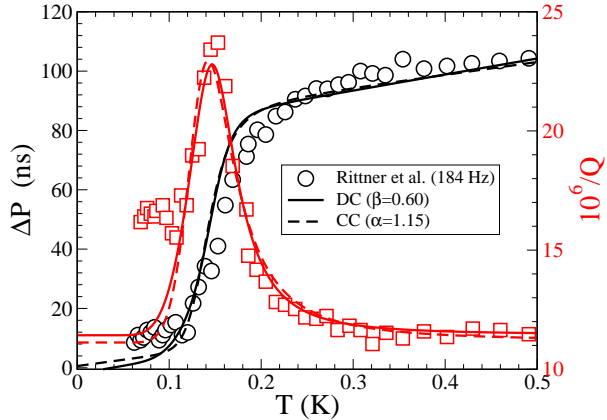}
\end{center}
\caption{(Color online) The period shift $\Delta P=P-P_0$
(black, left axis) and dissipation (red, right axis) vs.\ temperature
for solid \HeFour. The experimental data are from Rittner and Reppy, Fig.~2 of Ref.~\cite{Rittner06}.
A Davidson-Cole (DC) and Cole-Cole (CC) fit are shown. The DC fit was performed with parameters
$\beta = 0.60$,
$Q_\infty^{-1} = 11.4\cdot 10^{-6}$,
$f_0 = 184.2305$ Hz,
$\delta f = 69\ \mu{\rm Hz}$,
$A = 33.8$ mHz,
$\tau_0 = 0.439\ \mu{\rm s}$,
$D T_0 = 1.173$ K,
$T_0 = 0$ K,
$\alpha_T = 2.0\cdot 10^{-5}$ K$^{-1}$.
The CC fit used parameters
$\alpha = 1.15$,
$Q_\infty^{-1} = 11.1\cdot 10^{-6}$,
$f_0 = 184.2305$ Hz,
$\delta f = 20\ \mu{\rm Hz}$,
$A = 34.7$ mHz,
$\tau_0 = 1.95\ \mu{\rm s}$,
$D T_0 = 0.868$ K,
$T_0 = 0$ K,
$\alpha_T = 1.6\cdot 10^{-5}$ K$^{-1}$.
}\label{Fig1}
\end{figure}
\bigskip
\begin{figure}
\bigskip
\begin{center}
\includegraphics[width=0.70\linewidth,angle=0]{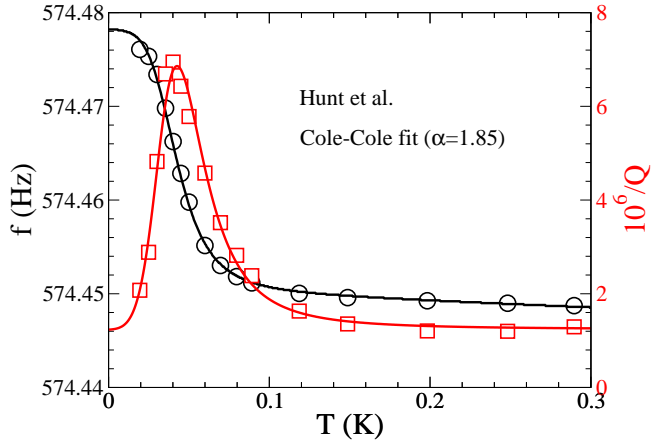}
\end{center}
\caption{(Color online) The resonant frequency (black, left axis) and dissipation (red, right axis) vs.\ temperature.
The experimental data
are from Hunt et al.\cite{Hunt09} with Cole-Cole (CC) parameters
$\alpha=1.85$,
$Q_\infty^{-1} = 1.23 \cdot 10^{-6}$,
$f_0 = 574.4768$ Hz,
$\delta f = 1.489$ mHz,
$A = 347$ mHz,
$\tau_0 = 2.52\ \mu{\rm s}$,
$D T_0 = 0.408$ K,
$T_0 = -44$ mK,
$\alpha_T = 2.43\cdot 10^{-5}$ K$^{-1}$.
}\label{Fig2}
\end{figure}
\begin{figure}[th]
\bigskip
\begin{center}
\includegraphics[width=0.70\linewidth,angle=0]{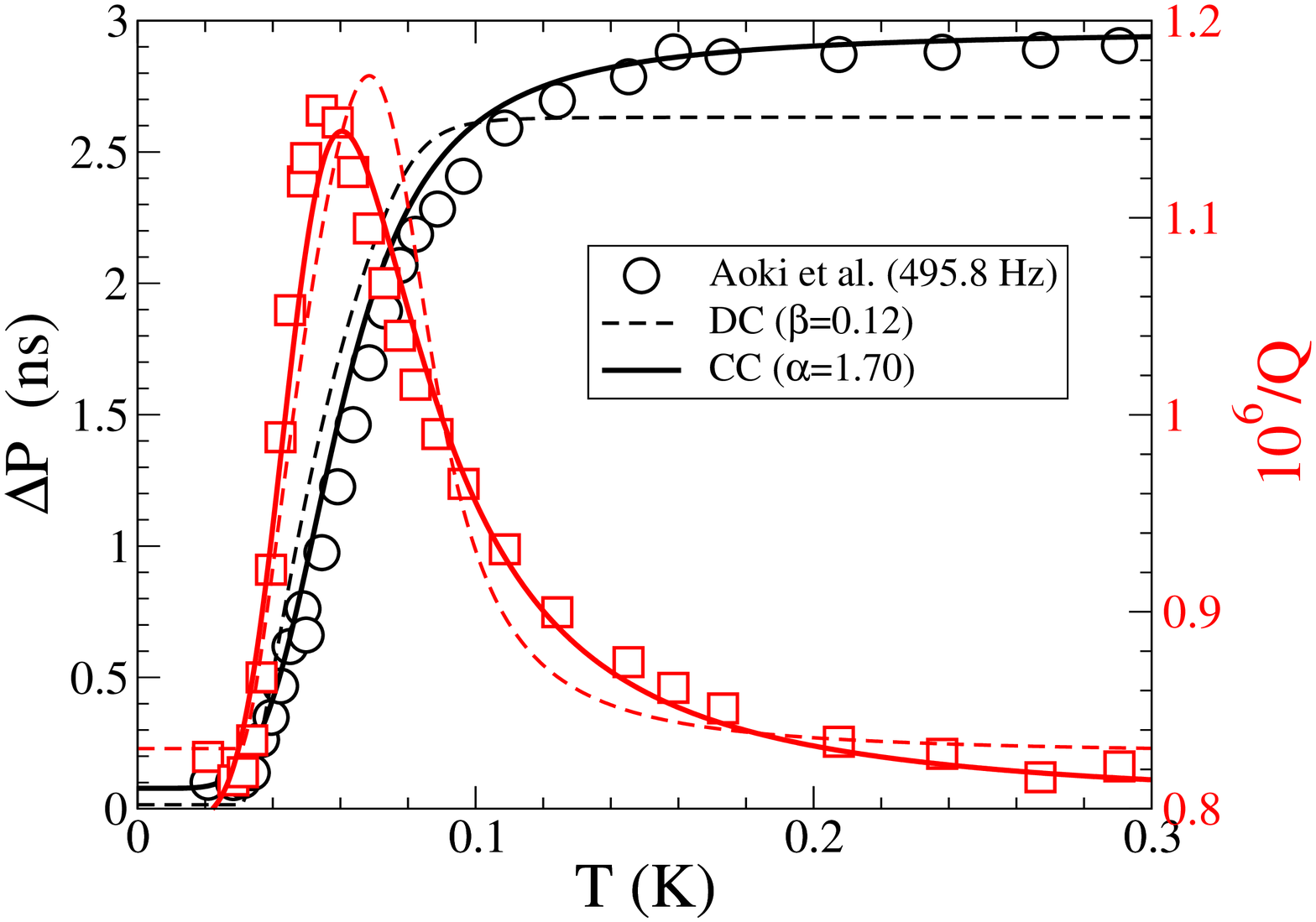}
\end{center}
\caption{(Color online) The period shift (black, left axis) and dissipation (red, right axis) vs.\ temperature.
The experimental data are from the in-phase mode (495.8 Hz) of the coupled double oscillator by
Aoki et al.\cite{Aoki07} The experimental data are already corrected for temperature dependence.
The DC parameters are
$\beta = 0.12$,
$Q_\infty^{-1} = 0.824\cdot 10^{-6}$,
$f_0 = 495.829$ Hz,
$\delta f=4.4\ \mu{\rm Hz}$,
$A = 8.19$ mHz,
$\tau_0 = 2.15\ \mu{\rm s}$,
$D T_0 = 0.306$ K,
$T_0 = 23.6$ mK,
$\alpha_T = 0$ K$^{-1}$.
The CC parameters are
$\alpha = 1.70$,
$Q_\infty^{-1} = 0.793\cdot 10^{-6}$,
$f_0 = 495.829$ Hz,
$\delta f=-19$ $\mu$Hz,
$A = 8.97$ mHz,
$\tau_0 = 13.2\ \mu{\rm s}$,
$D T_0 = 0.248$ K,
$T_0 = -17$ mK,
$\alpha_T = 0$ K$^{-1}$.
}\label{Fig3}
\end{figure}

All samples of solid \HeFour\ used in this study had in common that they were grown with the blocked capillary method
using commercial grade helium (\HeThree\ impurity level $\sim 0.3$ ppm).
Also, it is important to remember that both glass models (CC and DC) use only five fit parameters:
$g_0$, $\tau_0$, $D T_0$, $T_0$, and either an exponent $\alpha$ or $\beta$.
All other oscillator parameters are determined by normal state properties of the TO loaded with solid \HeFour.
In addition, we noticed during our analysis of the TO experiments
that in order to fit the glass models to the experimental data, we had to correct the resonant frequency by a small amount,
$f = f_0 + \delta f$, because in many reports $f_0$ is not available to
desired absolute accuracy or data are only reported relative to a high-temperature resonant frequency.
Furthermore, the fits were complicated by the experimental observation of a slight temperature dependence of the resonant
frequency at higher temperatures. To account for this drift in frequency of yet unknown origin, we approximated
$f^2_0(T) \approx f^2_0(0) [1-\alpha_T T]$ by a small linear-in-$T$ correction.

The TO experiment reported by Rittner and Reppy,\cite{Rittner06} see
Fig.~\ref{Fig1}, is in excellent agreement with the proposed
glass models. Both CC and DC glass distributions require exponents different from unity, which means that there
is a spread of relaxation times $\tau$.

In Fig.~\ref{Fig2}, we report an analysis of the measured data by Hunt et al.\cite{Hunt09} assuming a CC distribution of
relaxation times. As can be seen, an excellent
fit is obtained. For comparison, we also tried a DC distribution for relaxation times, but found only fair agreement.
It is worth pointing out that unlike in the Debye relaxation analysis by Hunt and coworkers
(a single overdamped mode), we do not require a supersolid component to simultaneously account for frequency
shift and concomitant dissipation peak.

Finally, in Fig.~\ref{Fig3}, we report a DC and CC analysis of the measured data by Aoki et al.\cite{Aoki07} for the
in-phase mode of their double resonance compound TO.
We obtain excellent agreement between experiment and glass model assuming a CC distribution,
while a DC distribution for glassy relaxation times results only in fair agreement.

\section{Conclusions}
To summarize, we have shown that a phenomenological glass model describing a small subsystem of solid
\HeFour\ can {\it simultaneously} account for the experimentally observed change in resonant period
(frequency) and the concomitant peak in dissipation.

Our analysis of TO experiments reveals that most are better described by a Cole-Cole
  distribution for glassy relaxation times.
Unlike for conventional structural or dielectric glasses, where the CC exponent $\alpha$ is usually less than unity,
we find consistently $\alpha > 1$. This may reflect on the possible nature of a quantum or superglass in solid helium.
Further, we derived a simple relation for the ratio of change in dissipation
and change in resonant frequency (period) that can explain the large ratios of order $\sim 10$ observed
in experiments.
The values for the glass exponents $\alpha$ or $\beta$ required to fit the experiments by the Rutgers and Cornell groups
point toward broad distributions of glassy relaxation times. This invalidates any attempt to describe these
experiments by a single overdamped mode (Debye relaxation).
These glassy relaxation processes
should also have significant effects on  thermodynamics and dynamics of solid \HeFour.
The key result of this work is that many TO experiments can be described assuming that a small
fraction of solid $^4$He undergoes a glass transition at low temperatures.
Whether or not there is a supersolid fraction present in solid $^4$He is beyond this analysis.
A frequency-tunable TO may differentiate between a glassy contribution leading to
an increase in the maximum frequency shift, $\Delta\omega_{max}\sim\omega_0 g_0/\alpha_{osc}$,
and no change in the dissipation shift, $\Delta Q^{-1}\sim g_0/\alpha_{osc}$,
with increasing $\omega_0$, while the frequency shift for a supersolid should decrease
with increasing $\omega_0$.
Our study shows that the unequivocal identification of supersolidity in  solid$^4$He is challenging and does require clear understanding of normal state dynamics.
Clearly, more dynamic studies probing the frequency or time response to a stimulus and detailed bulk characterization of samples
are necessary to investigate the differences between small subsystems of glassy, supersolid or superglassy origin.

\begin{acknowledgements}
This work was partially supported by the by the US Dept.\ of Energy at Los Alamos National Laboratory
under contract No.~DE-AC52-06NA25396 and by the Center for Materials
Innovation (CMI) of Washington University, St.\ Louis.

We are grateful to A.T. Dorsey, S.E. Korshunov, J. Beamish, J.M. Goodkind, H. Kojima and J.C. Davis for many stimulating discussions
on this topic.
\end{acknowledgements}


\end{document}